# Microwave Photocurrent from the Helical Edge state of InAs/GaInSb Bilayers

Jie Zhang[1*], Tingxin, Li[1], Rui-Rui Du[1,2*], Gerard Sullivan[3]

[1]*Department of Physics and Astronomy, Rice University, Houston, Texas 77251, USA*

[2]*International Center for Quantum Material, Peking University, Beijing 100871, China*

[3]*Teledyne Scientific and Imaging, Thousand Oaks, California 91630, USA*

**Abstract**

We measure microwave photocurrent in devices made from InAs/GaInSb bilayers where both insulating bulk state and conducting edge state were observed in the inverted-band regime, consistent with the theoretical prediction for a quantum spin Hall (QSH) insulator. It has been theoretically proposed that microwave photocurrent could be a unique probe in studying the properties of QSH edge states. To distinguish possible photoresponse between bulk state and helical edge state, we prepare Hall bar and Corbino disk from the same wafer. Results show that the Corbino disk samples have a negligible photocurrent in the bulk gap while clear photocurrent signals from the Hall bar samples are observed. This finding suggests that the photocurrent may carry information concerning the electronic properties of the edge states.

PACS#

78.20.Ls

78.67.-n

*jz38@rice.edu

*rrd@rice.edu

The helical edge state attracts special attention due to unique properties such as its robustness against backscattering, transport conductivity quantization and possible spin-momentum locking based on its topological nature [1-3]. Inverted InAs/GaSb double quantum wells (DQWs) [4-8] and InAs/GaInSb DQWs [9-11] have been demonstrated to be a favored platform to study the helical edge state properties. Due to the spatial overlap between the electron and hole wavefunctions, a hybridization gap is opened in the bulk along with conducting edges. By virtue of the front/back gates, Fermi level in the device can be continuously tuned [4-11]. Application of strain in InAs/GaInSb DQWs increases the bulk hybridization gap [9,10], thus allowing us to explore the nature of edge state without interference from the bulk.

Terahertz (THz) and infrared (IR) [12-15] spectroscopies have been powerful tools to probe 2D and 3D topological materials. Recently, the photovoltaic effect was used to study electron spin imbalance on the edges in HgTe QWs induced by circular photogalvanic effect in unbiased devices by selectively exciting spins [13] or utilizing the chiral properties of unique matter such as Weyl semimetals [14]. We report the observation of bias-current-polarized microwave (MW) photocurrent in strained-layer InAs/GaSb DQWs. In contrast to the work by G.C. Dyer et al. [12] with prevalent bulk photoconduction, we directly demonstrate that this signal comes from the edge states by performing the same measurement on a Corbino disk where edge state is shunted by metal electrodes. We measure the frequency and temperature dependence of the photoresponse and find that the application of a magnetic field can increase the photocurrent near the charge neutral point (CNP). This work investigates the optical and electric natures of pure edge state and lays the foundation of manipulating edge physics in this system with electromagnetic waves.

The experiments have been carried out on strained InAs/Ga$_{0.68}$In$_{0.32}$Sb DQWs grown by molecular beam epitaxy on n-doped GaSb substrates. The electron/hole QWs have well widths of 8 nm/4 nm respectively (Fig. 1(a)). Wafer details can be found in [10,11]. Typical Hall bar geometry with a length of 75 μm and width of 25 μm is patterned using photolithography as shown in Fig. 1(d). Ohmic contacts are made by selectively etching the samples down to the InAs QW and depositing Ti/Au as electrodes without annealing. A back gate is used to tune the Fermi level in the device. The ability to reach the quantum spin Hall (QSH) phase without a front gate enables the MW to interact with the helical edge states. The sample is immersed in He$^3$ liquid in a cryostat with a base temperature of 300 mK and equipped with a superconducting magnetic coil up to 8 T. The MW is produced by an Anristu signal generator which has a bandwidth of 2-40 GHz and is transmitted via a semi-rigid coaxial cable with an antenna in the end. The magneto-transport is measured by a standard low frequency lock-in amplifier which also provides the amplitude modulation of the MW in the photocurrent probing. A sufficient constant DC bias current (~10 nA) is used to provide directional selection for the excited carriers. In order to rule out the possibility that the observed photocurrent comes from the bulk contribution, a Corbino sample is fabricated on the same wafer using the same procedures, as shown in Fig. 1(b). DC transport measurement in Fig. 1(c) shows that when Fermi level is tuned in the bulk gap, the resistance of Hall bar sample has a maximum near the CNP while the conductance of the Corbino sample experiences a minimum indicating negligible bulk contribution. Photocurrent is measured by the photovoltage drop across a load resistor with a resistance much smaller than the resistance of the sample itself.

InAs/GaSb DQWs can be modeled by the Bernevig-Hughes-Zhang (BHZ) effective Hamiltonian [4]:

$$H = H_0 + H_{BIA} + H_{SIA} \qquad (1)$$

with bulk inversion asymmetry (BIA) term $H_{BIA}$ and structural inversion asymmetry (SIA) $H_{SIA}$ in addition to a quantitative hybridization correction. Projected to the effective 4-band basis, $H_{SIA}$ could be recognized as a k-linear Rashba term (neglecting the high-order k-cubic term) responsible for deviation from spin perfect alignment away from the CNP [16]. Despite these factors, QSH effect still persists in this system with the help of gate tuning, and resistance quantization can be observed in samples with size smaller than the phase coherent length [7,8,10,11]. For longer samples, the longitudinal resistance is larger than the quantizaton value [7-11]. The increasing resistance under magnetic fields near the CNP (Fig.2 (c,d)) indicates backscattering enhancement of the edge state since the contribution of the bulk state can be excluded. This can be interpreted as the destroying QSH states by breaking time reversal symmetry (TRS) [10,11]. Here resistance saturation happens at a smaller field, which is consistent with our previous studies [11] and a small excitation current is used to reduce heating. Theoretically large magnetic fields might also form Landau levels in the bulk ($B_\perp$) or shifting the subbands to the semi-metal regime ($B_{//}$). Neither of those effects is relevant to our case here [11]. Except from the truly DC insulating bulk (Fig. 2(a)), we also demonstrate in Fig. 2(b) that when excited by MW radiation, the bulk still contributes no photoconduction up to 8 T. Whereas, in the Hall bar sample data shown in Fig. 3 and Fig. 4, non-trivial photocurrent and photovoltage are observed. Therefore, it is reasonable to conclude that the photoresponse signal comes from the sample edges when the Fermi level is tuned in the bulk gap.

Generally when a semiconductor device is illuminated with MW radiation, charge carriers below the Fermi level absorb photon energy and become mobile within a certain relaxation time. Note in the MW photoexcitation scenario, inter-band resonant absorption is not relevant due to energy scale. These mobile charge carriers will drift under a bias electric field induced by external current application. Therefore, the photocurrent formed always has the same polarity with the external bias (Fig. 4(a)). However, the voltage drop caused by this mechanism is due to the fact that electrons and holes are spatially separated, meaning the voltage drop always tries to compensate for the external field. As a result, when a negative bias current is applied, the current shows a negative sign while the voltage is positive (Fig. 4(b)). Note that this case is different from photocurrents excited by circularly polarized radiation in unbiased devices in [13] where spin is selectively excited. It was recently proposed in [16] that unlike in the centro-symmetry model where only magneto-dipole interaction is able to couple the spin up and spin down branches; for a crystal lack of inversion symmetry, optical transition occurs due to electron-dipole interaction which is several orders of magnitude stronger. In the external-current-dependence measurement, a negligible zero-bias photocurrent shows the dichroism for circularly polarized light is small for this material.

Optical transition rate can be estimated by first-order perturbation theory where excited states are

$$|m\rangle \approx |m^0\rangle + \sum_k \frac{H'_{fm}}{E_m^{(0)} - E_k^{(0)}} |k^0\rangle. \quad (2)$$

Here $|m^0\rangle/|m\rangle$ is the initial/final state, $H'_{fm}$ is the off-diagonal matrix element for perturbation, $E_k^{(0)}$ is the energy from solving the non-perturbed equation. The resultant photocurrent is determined by the number of excited carriers and their drift velocity in an external electric field. Given the same microwave power and frequency, transition rate is set by the density of states (DOS) and perturbation strength $H'_{fm}$ of the initial/final state assuming certain interaction efficiency between electrons and photons. The electric field in the sample equals the bias voltage divided by the distance across the two leads. In the bulk gap where bulk contribution can be neglected (Fig. 2(a)), the external voltage is effectively applied on the edge state carrier. Therefore at the resistance peak shown in Fig. 2(c), the effective electric field also has a maximum. However, we observed a photocurrent minimum near the CNP (Fig. 3) which can only be explained via a reduction in transition rate. This is a reasonable result because when the Fermi level is positioned in the conduction and valence band, there is only one mode available: the linearly dispersed edge state. This results in a reduction of DOS for both initial and final state since the microwave photons have energy level of a few tens of $\mu eV$ which is much smaller than the bulk gap. Also note that due to the helical nature of the edge states, two states at a fixed $k$ have partially opposite spin (taking the $H_{BIA}$ and $H_{SIA}$ terms into account) [15], so the transition rate between two states with the same momentum is further reduced because of a reduction in $H'_{fm}$. Momentum conservation is another reason to constrain this transition because the momentum transferred in an electron-photon interaction can usually be neglected. Here data are collected by repeatedly downwards sweeping gate voltage to rule out temporal drift from the charge relaxation.

When we increase the microwave power, data in Fig.3 (a) shows that photocurrent increases with MW radiation until heating comes into play where thermal energy might also start to excite carriers. Note that 10 GHz microwave photons have equivalent thermal energy of about 0.5 K. In addition, we tune MW power to fix the temperature at 650 mK in order to rule out the attenuation effect in the circuit and vary MW frequency. Except from frequencies below 5 GHz where photon energy is much lower than thermal energy (heating effect dominates), no frequency dependence is observed up to 40 GHz (Fig. 3(b)). We attribute this phenomenon to the lack of clear absorption edge in the band structure in this system. Interestingly, there is a large shift (~0.1 V) of the photocurrent minima around three orders of magnitude larger than the MW photon energy which known mechanisms do not explain.

Theoretically, the application of magnetic field opens a gap on the edge state dispersion near the CNP [15,17]. However, the observation of this gap is extremely difficult in this disordered system where fluctuations provided by the randomly positioned scattering centers should be considered. For instance, the bulk inhomogeneities near the edge induced by the ionized dopant atoms in the QW, disordered nuclear spins [18] or the charge puddles [19,20] formed in the bulk. The resultant band structure might be locally shifted making the observation of the gap opening in the edge state by a magnetic field very difficult.

However, we find that for a fixed microwave frequency, the application of either perpendicular or parallel magnetic field increases both the photocurrent and photovoltage (Fig. 4) saturating at the value where DC resistance saturates as well. We attribute this phenomenon to the same TRS broken process that induces backscattering which may provide some momentum transferring in the electron-photon interaction. Note that the photocurrent signals also show saturation behaviors under a small magnetic field, similar to the measured resistance (Fig. 2(c), 2(d)). It is found that a perpendicular magnetic field has a larger effect than a parallel field (Fig. 4(a) inset) indicating a larger out-of-plane g factor. The same situation holds for radiation where the electric field is either along or perpendicular to the edge which is consistent with the particle-hole symmetry broken case in real semiconductor structures [16].

In conclusion, we have studied the photoresponse of edge states in InAs/GaInSb DQWs biased by external electric field under millimeter wave radiation. We demonstrate that the photocurrent purely comes from the edge state while bulk makes no contribution in the gap. The frequency, temperature, and magnetic field dependence of the photoresponse further explore the nature of edge state in this system, which we properly model with QSH physics.

*Acknowledgements* The work at Rice was supported by NSF Grant (No. DMR-1508644) and Welch Foundation Grant (No. C-1682).

**Figure Captions**

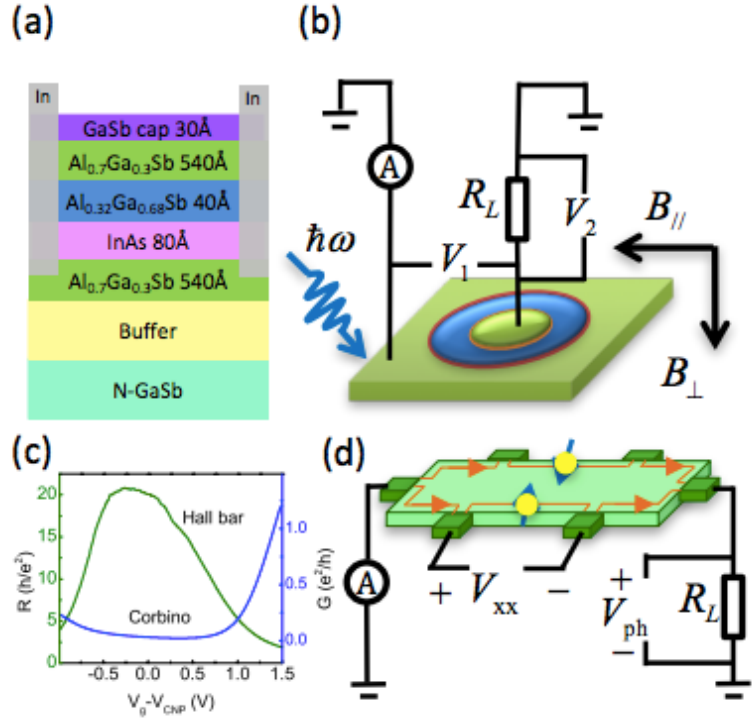

FIG. 1. (a) Structure of strain layer InAs/GaSb wafer used in the experiment, which hosts robust quantum spin Hall helical edge states. (b) Experimental setup of the photovoltage and photocurrent measurement on a typical Corbino disk with a backgate. (c) Resistance measured for Hall bar sample and conductance measured for the Corbino sample. (d) Photoresponse measurement setup on a Hall bar sample fabricated the same way with the Corbino sample shown in (b). Arrows along the edges illustrating spin-momentum locking and selection of current direction by DC bias current.

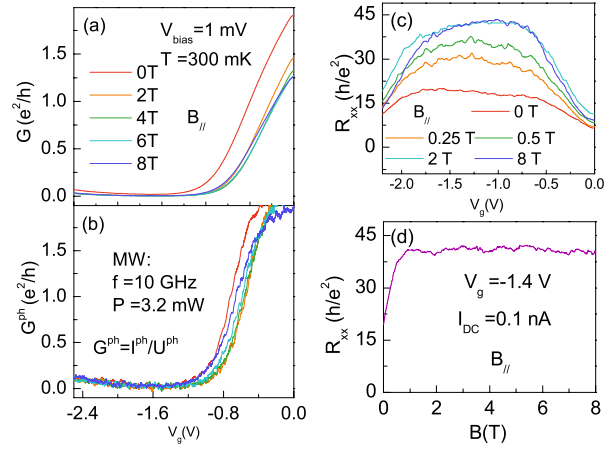

FIG. 2. (a) Magnetoconductance measurement of the insulating bulk state on a Corbino disk with a strong external parallel magnetic field (up to 8T). 2D bulk is accessed, sweeping the backgate voltage, which indicates a completely insulating behavior, giving no evidence of gap closing at increasing magnetic field. (b) Photoconductance measured on the same Corbino sample excited by a linearly polarized MW radiation. The MW amplitude is modulated by a lock-in amplifier. The miniscule photoconductance in the bulk gap demonstrates that the 2D bulk states make no contribution to the photocurrent. (c) Magnetoresistance measurements on a Hall bar device. The plateau increases with parallel magnetic fields, indicating an enhancement of backscattering under broken TRS. (d) Resistance of the Hall bar device in (c) measured by fixing the backgate at the plateau and sweeping a parallel magnetic field. A small bias current is used to avoid heating effect. Resistance doubled at 1T and showing saturation above 1T.

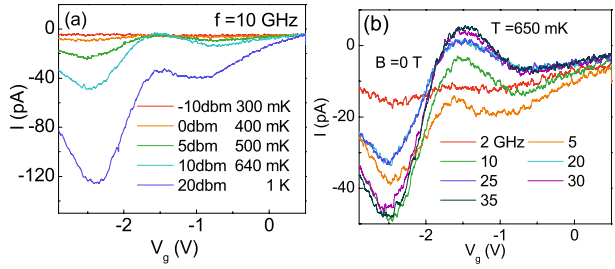

FIG. 3. (a) The MW power-dependent photocurrent traces measured in a Hall bar device with no magnetic field application and a large external bias current. Enhanced photocurrent on both sides of the gap boundary is detected with increasing MW radiation. Miniscule photocurrent at the CNP is maintained up to 10mW. This phenomenon starts to break down at 100 mW MW power due to heating. (b) MW frequency-dependent photocurrent measurement is shown by tuning the MW power to keep temperature at a fixed value of 650 mK. This is chosen instead of using a fixed MW input power in order to rule out the possibility of different attenuations at various frequency ranges. Shifting of the CNP towards lower backgate voltage is observed with increasing MW frequency.

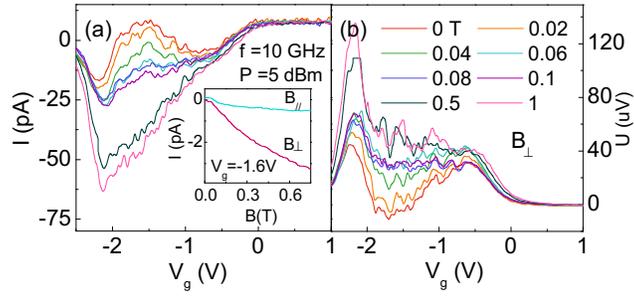

FIG. 4. Photocurrent (a) and photovoltage (b) measurement in a Hall bar device with fixed MW frequency and power. Backgate voltage is swept at different perpendicular magnetic fields. A moderate magnetic field is employed to break TRS, thus causing Rashba-type spin flipping between the two branches of helical edge states. This results in an enhancement of the photocurrent and photovoltage in the bulk gap. Comparison between the effect of parallel and perpendicular magnetic fields is displayed in the inset of (a) indicating a larger g factor for perpendicular fields.